# Magnetic measurements of Fermilab rapid-cycling Booster gradient magnets

J. DiMarco, D. Assell, T. Cummings, D. Johnson, V. Kashikhin,
M. Kifarkis, J. Kuharik, J. Larson, M. Mubarak, S. Poopathi, K. Triplett

*Abstract*— Fermilab is upgrading its Booster synchrotron to increase ramp rate and intensity. This is part of the Proton Improvement Plan (PIP-II) that will allow the Main Injector to achieve proton beam power of 1.2 MW within the next few years. This upgrade includes running the 55-year-old Booster magnets at 20 Hz instead of the usual 15 Hz, and construction of some shorter and wider aperture versions of these combined-function gradient magnets. Magnetic measurements were performed to characterize the present 15 Hz AC performance, and then again with 20 Hz ramp cycle to ensure performance and compatibility in this new operating regime. A 3 m-long curved flat-coil was developed for these measurements using Printed Circuit Board (PCB) technology. The probe also has a separate 0.5 m-long body-field probe, allowing integral, body, and end fields to be measured across 100 mm of the magnet aperture. The sampling rate for these measurements during the AC cycle was 200 kHz, and field resolution was better than 0.01%. Details of the probe, measurements, and results are presented.

## I. INTRODUCTION

The PIP-II (Proton Improvement Plan II) project is an upgrade to Fermilab's accelerator complex that will create an intense high-energy neutrino beam for the Deep Underground Neutrino Experiment (DUNE), and also enable a wide range of other physics research. Part of this project involves installing a new 800 MeV system for beam injection into the Booster ring. The Booster's repetition rate will be increased from 15 Hz to 20 Hz, allowing for 50% more beam acceleration while minimizing power loss [1]. In order to ensure adequate performance of the aging magnets of the Booster at the higher repetition rate, as well as operational continuity, it is important to characterize the magnet performance both at the current 15 Hz, and also after the 20 Hz upgrade. To that end, a magnet measurement system comprised of 3 m-long integral probe, body field probe, and data acquisition system was developed. The system allows for measurement during the AC ramping cycle at 200 kHz, with field strength and uniformity measurements having 0.01% resolution. This paper describes the system and analysis, and presents results from the 15 Hz measurements, along with first data at 20 Hz.

## II. MAGNET TESTING AREA

The Booster accelerator uses combined function magnets, or gradient magnets, to steer and focus the beam. The electrical circuit that powers the gradient magnets in the Booster consists of 48 resonant cells connected in series. Each cell has two Booster Gradient Magnets (one focusing BGMF and one defocusing BGMD), a choke, and three banks of capacitors mounted on a steel girder. The test stand described replicates one resonant cell of the Booster magnet circuit (Fig. 1).

The magnets positioned on top of the girder are original Booster spare magnets. The BGMF and BGMD have an inductance of ~11mH @1KHz. Capacitance for the resonant circuit is provided by three banks of capacitors positioned in the lower part of the girder. For 15 Hz operation, there are 17 total GE 16L555SJ4 capacitors with a total capacitance of 7590 µF.

The power supply is a PEI SR1039 500KW DC Power Supply driven by a 15 Hz or 20 Hz sine wave generated by a CAMAC 473 ramp card. The power supply is current regulated and is controlled via Fermilab's ACNET control system through an Allen Bradley Micro Logix PLC. A high current low pass filter system is wired with the girder to remove undesired ripple from the power supply output.

Cooling for the magnets, choke, and power supply is provided by the Remnant Low Conductivity Water system (LCW). The LCW temperature and pressure of the test stand are made to match the Booster LCW system of approximately 90°F and 105 psi. Flow through each magnet is about 3 GPM.

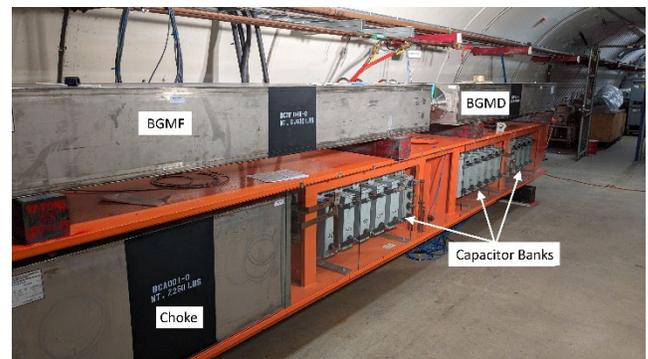

**Fig. 1.** Magnet test stand girder.

[1]Submitted for review August 4, 2025. This work was supported by the U.S. Department of Energy, Office of Science, Office of High Energy Physics. The authors are with Fermi National Accelerator Laboratory, P.O. Box 500, Batavia, IL 60510, USA, (corresponding author e-mail: dimarco@fnal.gov).

## III. MAGNETIC MEASUREMENTS SYSTEM

In order to capture the integral field at all times during the AC cycle, a stationary probe of 3.2m was developed, relying on inductive pick-up loops and the dB/dt from magnet current ramping to measure field change (Fig. 2).

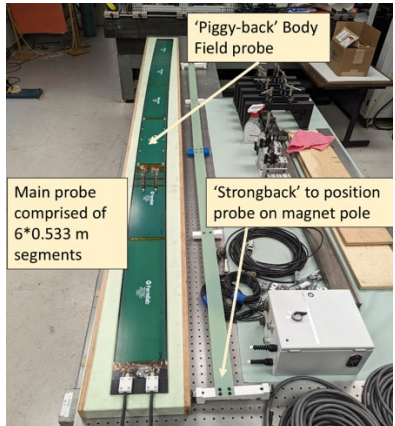

**Fig. 2.** Curved, 3.2 m probe and strongback.

A 535 mm-long curved PCB was designed that would match the arc of the magnet (not quite correct, because the magnets have shifted laminations rather than a true arc, but the difference is an offset of about 25 μm at the extremes, and so is negligible to the measurement at the 1e-4 level), and 6 of these boards connected end-to-end mounted on a G-10 platform comprise the probe.

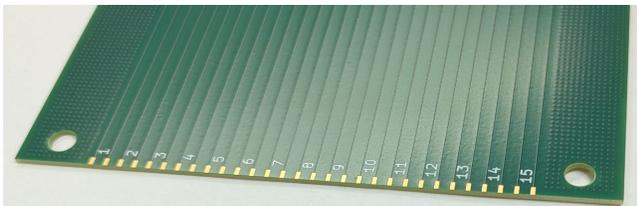

**Fig. 3.** End portion of 535 mm PCB probe section.

There are 15 windings on the PCB probes, each having 1 pair of traces and 3.5 mm width, with centers separated by 7.0 mm. The total aperture width being sampled therefore is 101.5 mm. The traces on the PCBs do not close, rather they end in solder pads at the end of the board, which can then either be continued by soldering to an adjacent PCB to extend the trace length, or if it is the end PCB, soldering a wire to bridge across the trace pair, thereby forming the loop (Fig. 3). The integral probe platform sits atop a 'strongback' which has 3D-printed pieces that match the contour of the pole, so that the strongback constrains the probe alignment with respect to the pole geometry both horizontally and vertically (Fig. 4). A spacer bar with 7mm thick edges, is used on the strongback to center the probe on the magnet pole, and removal of the spacer bar allows the probe to be shifted to stops which result in the probe being translated +/-7 mm from its central position, so that neighboring loops can then sample the same portion of field in the aperture – this is an important feature which allows for cross-calibration of the widths among the 15 loops. At the lead end of the probe, the pads are soldered to 27-pin Omnetics micro-connectors, and then are connected via 24-conductor, individually shielded, twisted-pair cables, to the data acquisition.

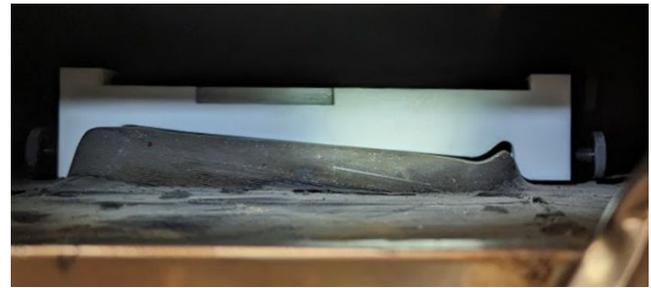

**Fig. 4.** Strongback positioned wrt pole via 3D-printed fitting.

## IV. ANALYSIS AND RESULTS

Most data presented are from the BGMD magnet, unless noted otherwise

### A. Integral Probe

Raw voltage data from the 15 loops of the probe are acquired by 24-bit 200 kHz simultaneous sampling Dewesoft ADC modules. In addition to the probe voltages, the current readout from the power supply transductor is also acquired. Signals timing is synched to the current cycle start via TTL control trigger. One hundred current cycles of data are acquired in a typical 'run' (about 6 seconds duration) and probe voltages are averaged and integrated to determine cumulative flux; linear drift is subsequently removed. Runs are performed with the probe centered, and for calibration purposes, horizontally offset by +/- 7mm so that neighboring loops measure the same portion of field. Note that small errors in the exact displacement are compensated for by scaling the averages across the overlap regions of windings in the displaced and centered positions (valid since the error in displacement applies to all windings). The local 'neighbor' comparison can then be propagated, starting with defining the center winding gain=1, and working towards each side of the probe, cumulatively multiplying the relative scalings. An example of the inter-calibration results of windings is shown in Fig. 5.

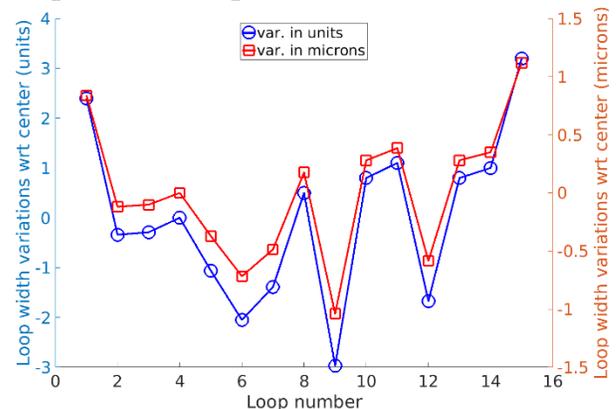

**Fig. 5.** Loop width variations across the integral probe (plotted in units on the LHS and microns on the RHS).

Corrections are on the order of 3 units owing to probe width variations (since the probe ends extend beyond the magnet end,

it is the widths, not lengths, that cause the differences), which correspond to about 1 micron in winding width.

During 15 Hz operation, the Booster cycle runs magnet current from 100 A to 1000 A. However, since the probe relies on the field change to measure flux, it is necessary to add the portion of field ramping from 0 to 100 A to the nominal cycle to determine actual field values. This is accomplished by making a separate run, ramping 0 to 1000 A, and adding the 0 to 100 A flux change as an offset to the 100 A – 1000 A cycle result.

A polynomial fit is performed at each data sample across the 15 loops to represent the field, $B_y$ as a function of $x$ position as

$$B_y = C_1 + C_2 * \left(\frac{x}{R}\right)^1 + C_3 * \left(\frac{x}{R}\right)^2 + \cdots + C_n * \left(\frac{x}{R}\right)^{n-1}, \quad (1)$$

where R is a reference radius (25.4 mm), $n$ the order, and the $C$s are the fit coefficients determined by solving the matrix expression

$$\begin{pmatrix} (K_1(w_1)) & (K_2(w_1)) & (K_3(w_1)) & & (K_9(w_1)) \\ (K_1(w_2)) & (K_2(w_2)) & (K_3(w_2)) & \cdots & (K_9(w_2)) \\ (K_1(w_3)) & (K_2(w_3)) & (K_3(w_3)) & & (K_9(w_3)) \\ & \vdots & & \ddots & \vdots \\ & \vdots & & & \vdots \\ (K_1(w_{15})) & (K_2(w_{15})) & (K_3(w_{15})) & & (K_9(w_{15})) \end{pmatrix} \begin{pmatrix} C_1 \\ C_2 \\ C_3 \\ \vdots \\ \vdots \\ C_9 \end{pmatrix} = \begin{pmatrix} \varphi_1 \\ \varphi_2 \\ \varphi_3 \\ \vdots \\ \vdots \\ \varphi_{15} \end{pmatrix}. \quad (2)$$

Here in (2), $\varphi$ are fluxes of the windings, and $Kn$ are defined by

$$K_n(w) = \frac{LR}{n} * \left(\left(\frac{w^+}{R}\right)^{n-1} - \left(\frac{w^-}{R}\right)^{n-1}\right), \quad (3)$$

where $w^+$ and $w^-$ are the x-positions of the two traces of a given winding and L is the length (=1 for the integral probe). The field shape residual at max. current with dipole and gradient suppressed is shown in Fig. 6.

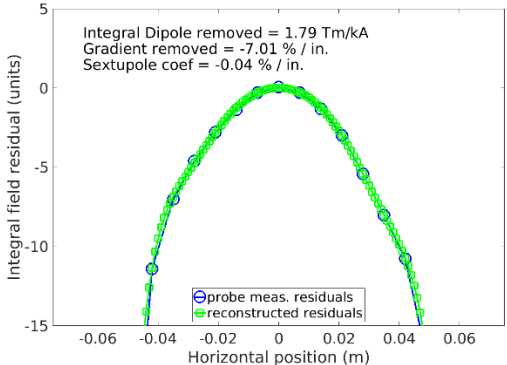

**Fig. 6.** Integral field residual at 997A during AC ramp cycle from 0.1 kA to 1 kA for defocusing magnet BGMD051.

The field shape residual across the full aperture shows little change as a function of current (Fig. 7).

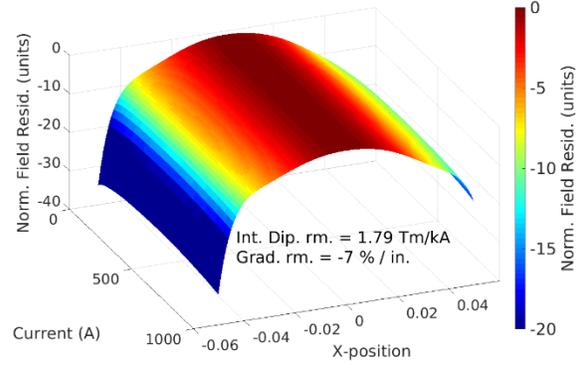

**Fig. 7.** Integral field residual over the full AC cycle.

After compensating for gain and offset of the current signal, the hysteresis of the magnet transfer function is shown to be ~0.1 %, during the AC cycle. A comparison of this hysteresis for 15 Hz and 20 Hz 100-1000 A cycles is shown in Fig. 8 – the hysteresis behavior at the new frequency is very close to the original.

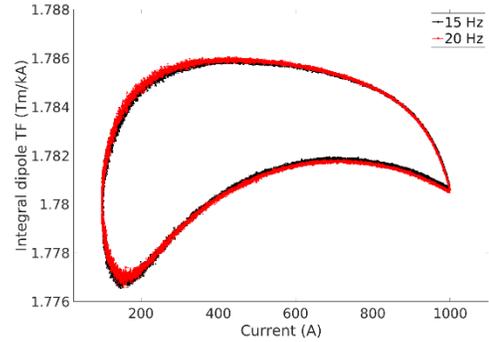

**Fig. 8.** Integral transfer function (TF) vs current at 15 Hz and 20Hz cycle operation for BGMD051.

The integral field shape residual for the BGMF magnet is shown at max. current in Fig. 9, and the integral TF hysteresis at 15 Hz and 20 Hz is shown in Fig. 10.

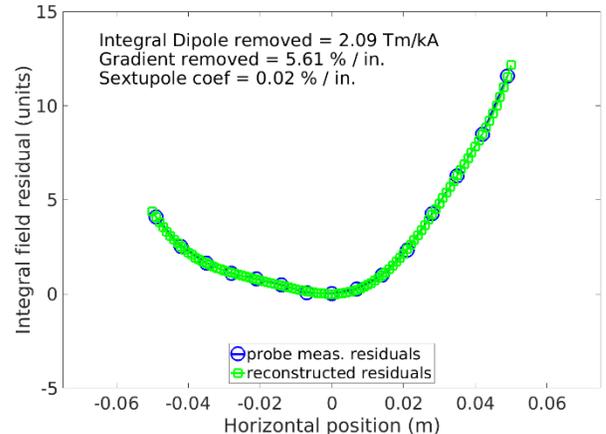

**Fig. 9.** Integral field residual at 999A during ramp cycle from 0.1 kA to 1 kA for focusing magnet BGMF046.

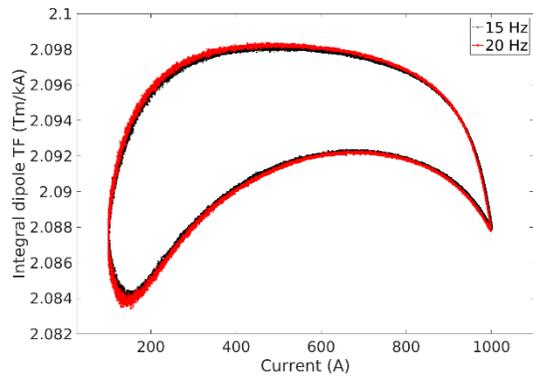

**Fig. 10.** Integral transfer function (TF) vs current at 15 Hz and 20Hz cycle operation for BGMF046.

*F. Body/end measurements*

A single one of the 535 mm PCBs was used as a separate 'short probe' to measure the body field, and to map the end field region. In order to position the probe, a separate strongback with contoured 3D-printed pieces to interface to the magnet pole was constructed. Here the probe was centered by means of G10 wheels which constrained the probe laterally but allowed axial motion. The PCB was then connected to a Zaber linear stage with 150 mm stroke via a bracket and arm, allowing remote positioning from outside the powering enclosure (Fig. 11). Since the probe is curved, as it is pulled axially, its position relative to the pole remains correctly centered. The arm which connects the stage to the probe has two pivots which allows this curved trajectory guided by the wheels, despite the linear pull. The wheels centering the probe are removable and can be replaced with ones which again enforce a 7mm shift either side from center for inter-calibration of the short probe windings. Positions for which the wiring at the end of the probe was within the magnetic field, showed calibrations where the added area from the soldering connections dominated the variation among windings. For the position furthest within the magnet (body field), the area variation was as large as 15 units (compared to 3 units for the integral probe (see Fig. 5)), which corresponds to about 0.75 mm in length. As the probe end was stepped outward, into the magnet end, the amount of field the connections saw decreased; as it went into zero field, the calibration among windings returned to the level observed in the integral probe (width error dominated).

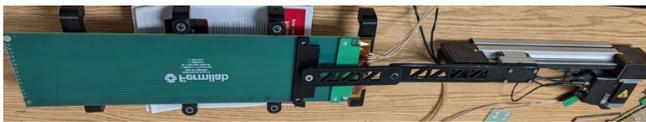

**Fig. 11.** Short probe and motion stage for measuring body and end fields of the magnets.

Body field results show some small differences compared to the integral (e.g. 3 unit sextupole instead of 4, but also 3 unit decapole instead of 1 unit), but were for the most part very similar. To measure the end field, runs were performed stepping the probe from the body field through the end in 5 mm intervals (30 positions). Individual inter-calibrations were determined and applied to the runs at each position. The change in going from one position to the next subtracts 5 mm of body field, as the probe end inside the magnet moves out, but then adds the 5 mm of field contribution from the probe in the end field. Subtracting these adjacent measurements and adding back 5mm worth (i.e. 5/535) of the body field, therefore gives the local field in the 5 mm portion of the end. Performing this for all consecutive positions gives the magnet field strength in the end as shown in Fig. 12 (note that it was necessary to perform a second scan with the stage shifted by 40mm to have enough range).

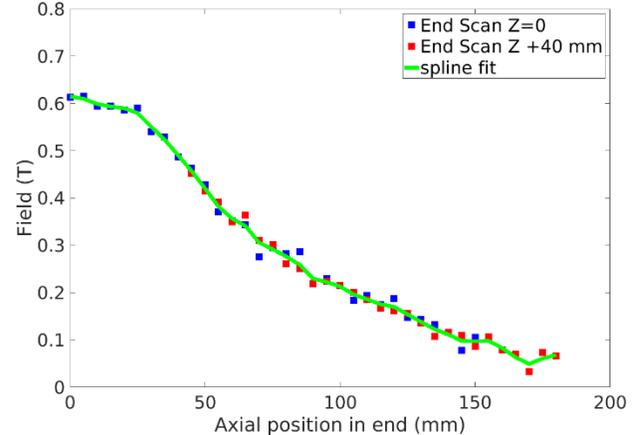

**Fig. 12.** Local end field as a function of position through the end of the BGMD magnet.

The field strength profiles across the aperture for Z-positions through the end are shown in Fig. 13 at 1000A. As for all results, it is similarly available at any given current in the cycle.

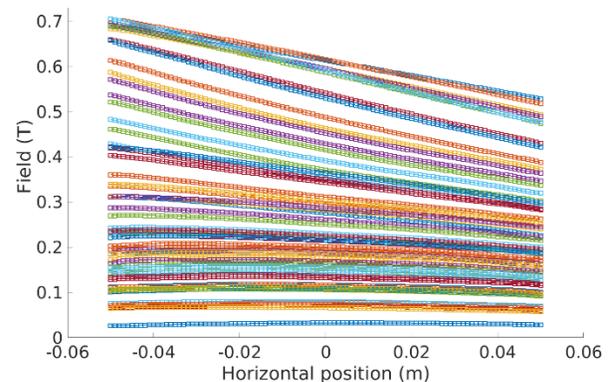

**Fig. 13.** Local fields across the aperture at 1000A, shown for axial positions stepping from the body field through the end (top curves to bottom, respectively).

V. CONCLUSION

A measurement system has been developed for the AC measurements of the Booster gradient magnets at Fermilab. The system features inter-calibrations of the inductive pick-up loops at the level less than 0.5 micron, allowing resolution of 1 unit of field during the AC ramp while sampling the field at 200 kHz. A similar short probe has also been used to measure the body and end field, and characterization of the magnets at 15 Hz AC current is complete. First 20 Hz results indicate hysteresis and field residual profile very similar to the original 15 Hz results.